# Intrinsic Resolution of Compton Electrons in CeBr$_3$ Scintillator using Compact CCT

Snigdha Sharma, V. Ranga, S. Rawat, M. Dhibar, *Student Member, IEEE,* G. Anil Kumar, *Member, IEEE*

*Abstract*—CeBr$_3$ is emerging as one of the best scintillators having properties almost similar to Cerium doped lanthanum halide scintillators. We have measured, for the first time, the intrinsic energy resolution of Compton electrons in a cylindrical 1″ × 1″ CeBr$_3$ detector using the sources, namely, $^{137}$Cs, $^{22}$Na and $^{60}$Co employing Compton Coincidence Technique (CCT). We have used PIXIE-4 data acquisition system which makes the measurement setup quite compact. The results have shown that non-proportionality is the major factor in limiting the overall energy resolution of CeBr$_3$ and the intrinsic resolution in CeBr$_3$ arises due to processes other than the scattering of electrons inside the scintillator. We have also studied the dependence of intrinsic energy resolution on the coincidence window and optimized its value for a given source.

## I. INTRODUCTION

Among Cerium halide scintillators, CeBr$_3$ is proved to be promising scintillator in recent times [1]. The great importance of these detectors is associated with their properties such as good energy resolution (~4% at 662 keV), excellent timing resolution (~93ps at 511 keV), high light output (~68000 photons/MeV), high stability in light output during temperature changes, high effective atomic density (~5.2 g/cm$^3$), possibility for growing in large volume, etc. [2]-[7]. Another attractive property of CeBr$_3$ is its higher radio-purity when compared to Lanthanum halide scintillators. The small amount of internal radioactivity in CeBr$_3$ is mainly due to the $^{227}$Ac contamination in the raw material of the detector. The overall internal radioactivity of CeBr$_3$ was reported to be 7-8 times less than that in LaBr$_3$:Ce and LaCl$_3$:Ce detectors, making CeBr$_3$ more sensitive to low count rate applications [8]. All these properties are responsible for its application in well logging devices [6], sub-nanosecond nuclear half-life using time-of-flight measurements [7], 3-D imaging in gamma ray astronomy [9], radioactive aerosol monitoring devices [10], Naturally Occurring Radioactive Materials (NORM) measurements [11], solar gamma-ray spectrometer GRIS [12], PING-M experiment to investigate solar X-ray activity [13], remote sensing applications [14], space missions [15],[16], water treatment facilities [17], steelworks [18], studying fast ions at JET [19], high energy gamma spectroscopy [20], etc. A huge amount of testing and characterization work is in progress by several groups to fully understand the limitations and potentials of these crystals. The present work aims to understand the factors affecting the energy resolution of CeBr$_3$ detectors.

Energy resolution is one of the important parameters which greatly affects the productivity of any detector. Finding out the ways to improve the resolution of detectors are still in progress. It is well established that the major limitation to the overall energy resolution of a detector is the intrinsic resolution, which arises due to non-proportionality of light output of the crystal by means of scattering of electrons (δ-rays) and landau fluctuations [21]. The origin of intrinsic resolution, however, is an open problem and yet to be fully understood [22].

In order to measure the intrinsic energy resolution of a detector, a Compton spectrometer based method was first proposed, in 1994, by Valentine and Rooney [23]. The method, benchmarked as Compton Coincidence Technique (CCT) in the year 1996, provides accurate characterization of light yield non-proportionality with the measurement of electron response of the detector [24]. Since then, CCT has been widely used to determine the non-proportionality in scintillators [25]-[33]. The technique is based on the detection of Compton scattered γ-rays and the basic principle of the method is to register in coincidence the signals from scattering of γ-ray inside the tested detector followed by absorption of the scattered ray inside the reference detector. Significant number of experimental studies were done by Swiderski *et al.* [34]-[38] to determine the non-proportionality and intrinsic resolution of Compton electrons in LaBr$_3$:Ce, LYSO:Ce, CaF$_2$:Eu, BC408, EJ301, NaI:Tl, CsI:Tl, CsI:Na and Xe gas detectors. In case of CeBr$_3$, although few detailed studies on the non-proportionality response of CeBr$_3$ and ways to improve it are available in the literature [2], [39]-[41], no experimental measurements on intrinsic resolution of Compton electrons in CeBr$_3$ detectors were reported in the literature. In this paper, we report our studies on intrinsic energy resolution of Compton electrons in CeBr$_3$ detectors by employing CCT. The measurements of Compton electrons were done using a PIXIE-4 multi-channel digital gamma processor

Manuscript submitted on June 24, 2017.
"This work was supported in part by the DST Govt. of India under fast track grant SR/FTP/PS-032/2011".

The authors are with the Radiation Detectors and Spectroscopy lab, Department of Physics, Indian Institute of Technology Roorkee, Roorkee-247667, Uttarakhand, India (e-mail: himasnig@gmail.com, 1viru1994@gmail.com, rawatsheetal1991@gmail.com, physics.monalisha3@gmail.com, anilgfph@iitr.ac.in).

which makes whole experimental setup relatively more compact. The optimization of the coincidence window value, which should be set for a gamma source of given activity, is also discussed.

## II. EXPERIMENTAL DETAILS

Fig.1 shows the schematic of the experimental setup used in the present work. The tested detector is a 1″×1″ CeBr$_3$ crystal optically coupled to a 2″ Hamamatsu R6231 PMT operated at +600 V and the reference detector is a 2.18″×2.36″ coaxial type reverse electrode HPGe detector biased with −4500 V. The detectors were kept face-to-face at a distance of 4 cm. A $^{137}$Cs source was placed between them. For processing the signals from detectors, we have used PIXIE-4 multichannel data acquisition system (supplied by XIA LLC) which provides digital spectrometry and waveform acquisition for four input signals per module [42]. Some of the main features of PIXIE-4 system include coincident data acquisition across channels and modules, pulse heights measured with up to 16 bits accuracy on each of the four channels, programmable gain, input offset, trigger and energy filter parameters, etc. It works with common resistive feedback preamplifiers of either signal polarity. Detailed technical information about PIXIE-4 system can be found in Ref. [43]. The present PIXIE-4 module with PCI-PXI based architect clearly meet the requirements of CCT resulting in a compact experimental setup for measuring the Compton electrons. The signals from both detectors through their respective preamplifiers were directly fed to channels 0 and 1 of PIXIE-4 module. The hit pattern of PIXIE-4 was so adjusted that only the signals detected in coincidence in channel 0 and channel 1 were recorded.

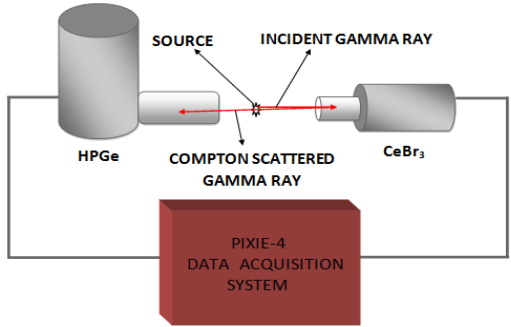

Fig. 1. Schematic of the experimental setup.

## III. DATA ANALYSIS

We have recorded the events registered in coincidence in both detectors using PIXIE-4 data acquisition system. Fig. 2 shows a typical spectrum of $^{137}$Cs source measured with HPGe detector in coincidence mode with a coincidence window of 1 $\mu$s. The spectrum shows prominent peak at Compton edge along with a photo peak and a backscattered peak. Similarly, Fig. 3 shows the spectrum of $^{137}$Cs source measured with CeBr$_3$ detector in coincidence mode. The photo peaks in Fig. 2 and Fig. 3 are due to the accidental coincidences of two gamma rays detected in both the detectors.

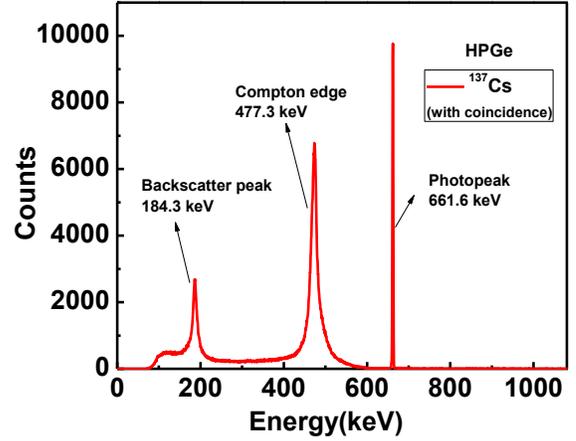

Fig. 2. Energy spectrum of $^{137}$Cs source measured with HPGe in coincidence with CeBr$_3$.

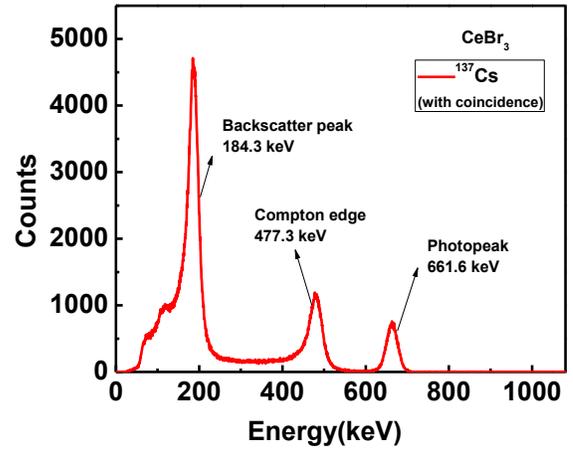

Fig. 2. Energy spectrum of $^{137}$Cs source measured with CeBr$_3$ in coincidence with HPGe.

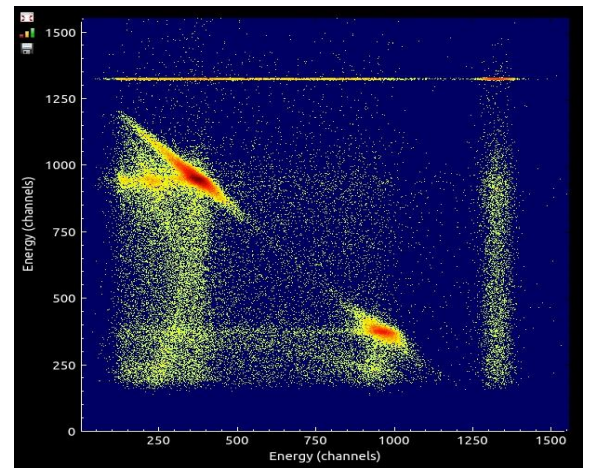

Fig. 3. 2D spectrum of coincident events in both HPGe (y-axis) and CeBr$_3$ (x-axis).

We are interested in only those events in which gammas are backscattered from CeBr$_3$ detector at an angle of 180° into the HPGe detector, thus giving off maximum energy to the electrons in CeBr$_3$. As $^{137}$Cs source emits gammas of 661.6 keV energy, the energies deposited in CeBr$_3$ and HPGe are 477.3 and 184.3 keV, respectively. Fig. 4 shows the 2-dimensional spectrum of the coincident events registered in both the detectors. The figure shows a vertical line in CeBr$_3$ (x-axis) and a horizontal line in HPGe (y-axis). These lines consists of the registered events corresponding to accidental coincidences of two gamma rays fully detected in both detectors. The figure also shows the events that give sum of energies equal to 661.6 keV. The measurement of energy resolution of Compton electrons in CeBr$_3$ detector demands the gating on the events that deposit energy in HPGe detector corresponding to the backscattered peak. The excellent energy resolution and precise calibration of HPGe detector enabled us to pick up the exact events of our interest. We have used RADWARE [44] software for offline gating. In order to understand the effect of the gate width or energy window width on the energy resolution, we have measured the energy resolution of Compton electrons for different values of gate width starting from 2 keV. Fig. 5 shows the plot of energy resolution of Compton electrons in CeBr$_3$ versus gate width in HPGe for $^{137}$Cs source. The increase in the gate width causes the widening of scattering angles resulting in increase in the FWHM of the peak. This leads to worsening of the energy resolution as evident from the figure. The gating was done on HPGe energy axis at 184.3 keV with a gate width of 2 keV and the projection was taken onto the CeBr$_3$ axis. The resulting gated spectrum of CeBr$_3$ is shown in Fig. 6. The figure clearly shows a Gaussian shaped peak at Compton edge corresponding to the energy of 477.3 keV. The resolution $\left(\frac{\Delta E}{E}\right)$ of Compton electrons can be easily determined from this peak. The intrinsic resolution ($\delta_{int}$) was calculated, neglecting the transfer component [25], by using the equation [27]:

$$\delta_{int} = \sqrt{(\frac{\Delta E}{E})^2 - \delta_{st}^2} \quad (1)$$

where $\delta_{st}$ denotes the photoelectron statistical contribution given by

$$\delta_{st} = 2.35 \times \sqrt{\frac{1+\varepsilon}{N_{phe}}} \quad (2)$$

where $N_{phe}$ is the number of photoelectrons and $\varepsilon$ is the gain variance of the PMT. For R6231 PMT, the gain variance was taken to be 0.28 [45]. $N_{phe}$ was calculated from the average value of absolute light yield reported for encapsulated sample of the same size [5]. The quantum efficiency of PMT was taken to be 30% as provided by the manufacturer [46].

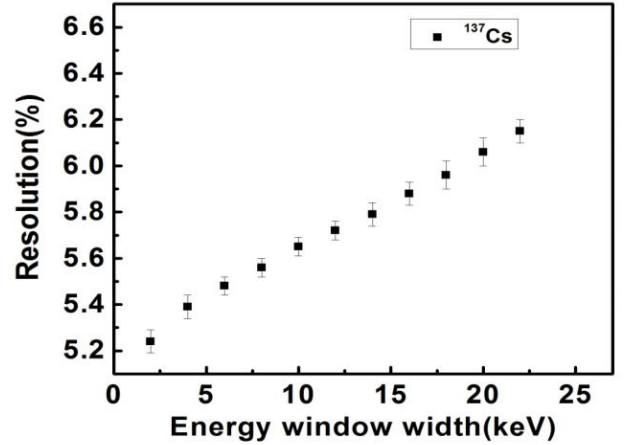

Fig. 5. Resolution of Compton electrons in CeBr$_3$ versus energy window width in HPGe for $^{137}$Cs source.

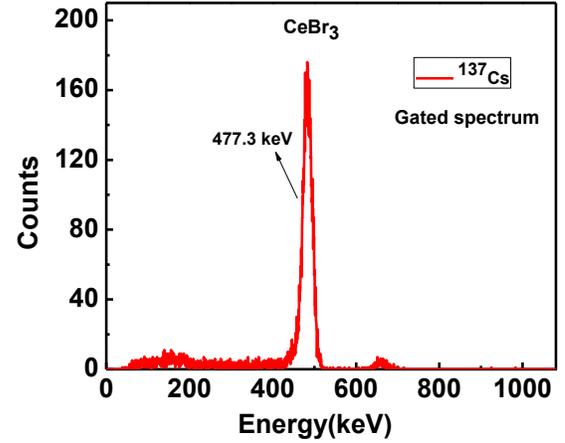

Fig. 6. Projected spectrum of CeBr$_3$ with HPGe gated on backscattered peak

## IV. RESULTS AND DISCUSSION

### A. Intrinsic resolution

In order to understand the effect of intrinsic resolution on overall energy resolution of CeBr$_3$ detector, we have estimated the intrinsic resolution of Compton electrons in CeBr$_3$ using Equation (1) considering the sources $^{137}$Cs, $^{60}$Co and $^{22}$Na. For $^{60}$Co, the intrinsic resolution of both 963.4 keV and 1118.1 keV Compton electrons were calculated, corresponding to gamma energies 1173 and 1332 keV respectively. For $^{22}$Na, in order to reduce the counts of directly detected 511 gammas, the source was slightly shifted from the original position. So, HPGe was gated on 184.8 keV, corresponding to the scattering angle of 140° from CeBr$_3$, giving rise to a Compton peak at 326.3 keV. The results are presented in Fig. 7. The figure also shows the intrinsic resolution of two other types of events. One, full energy peaks due to gamma rays backscattered in HPGe and then absorbed in CeBr$_3$. These peaks corresponds to the Compton electron energies, namely, 184.3 keV, 209.8 keV and 214.4 keV. Two, full energy peaks due to gamma rays coming directly from the source and deposited full energy in CeBr$_3$. The data presented in Fig. 7 clearly confirms the larger contribution of non-proportionality component in overall energy resolution of $CeBr_3$ detector, as already reported by Quarati *et al*. [2]. The difference between intrinsic resolution of Compton events and photo peak events of almost similar energy is clearly evident. For example, the intrinsic resolution of Compton electrons of energy 1118 keV is 2.2% and of full energy peak corresponding to 1173 keV is 3.5%. Similarly, the intrinsic resolution of Compton electrons of energy 477 keV is 4.1% and of full energy peak corresponding to 511 keV is 5.3%. This confirms that intrinsic resolution not only arises due to the scattering of electrons inside the scintillator but also due to the other processes such as self-absorption and re-emission processes [40]. Measurements with different sizes of $CeBr_3$ are required to better understand the factors contributing to the non-proportionality in these scintillators.

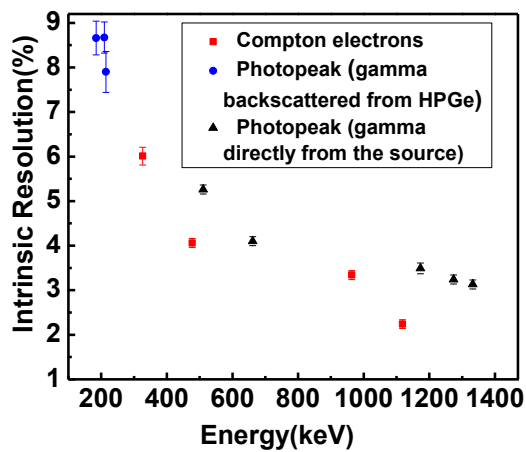

Fig. 7. Intrinsic resolution for $CeBr_3$

### B. Optimization of the coincidence window value

From the activity of the $^{137}Cs$ source, we could estimate the time difference between the emissions of two successive gamma rays from the source. It was calculated to be around 21 $\mu$s. If the coincidence window value is set above 21 $\mu$s then there would be an increased number of accidental coincidences due to more number of gamma rays being emitted during the time interval. Also, some of the actual coincident events would be lost. For example, if a gamma scatters at an angle in HPGe, depositing some energy, the acquisition system will wait for a coincident event in $CeBr_3$. If the coincidence window value is large enough, another gamma may enter $CeBr_3$ depositing the required energy of 477.6 keV and backscattering to HPGe, deposing an energy of 184.3 keV. The system will consider the first two events as coincident ones rather than the last two, which are the coincident events we actually require. To avoid this, window value should be set well below the time required for the emission of two or more gamma rays. To prove this statement we extended our work and recorded the data for 870 $\mu$s which is the maximum coincidence window value possible to set in PIXIE-4. We get a better understanding when we compare the gated spectrum for coincidence window values ranging from 60ns to 870$\mu$s as shown in Fig. 8. Clearly, in order to get a well pronounced Compton edge peak with considerable number of counts, one should select a coincidence window value which is neither too high nor too low and is well below the time required for the emission of two or more gamma rays.

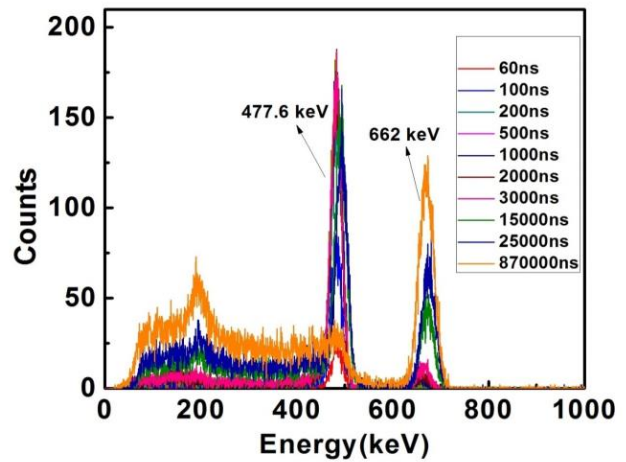

Fig. 8. Comparison of gated energy spectrum of $CeBr_3$ for coincidence windows ranging from 60 *n*s to 870 $\mu s$

### V. SUMMARY AND CONCLUSION

$CeBr_3$ scintillators are proved to have advantages over equivalently sized $LaBr_3$:Ce detectors. We could successfully employ Compton Coincidence Technique in order to measure the intrinsic resolution of Compton electrons in $CeBr_3$ of volume 12.87 cm$^3$. The use of PIXIE-4 makes the complete measurement setup quite compact. The results clearly suggested that the intrinsic resolution in $CeBr_3$ arises from processes other than the scattering of electrons inside the scintillator. Also, an optimization of the coincidence window value has been done based on the activity of the source in order to get the best possible Compton edge peak. As the effect of non-proportionality of light output on the energy resolution is more prominent for large volume crystals, further work is in progress to study Compton electrons in large volume $CeBr_3$ detectors which have recently become commercially available.

ACKNOWLEDGMENT

Authors would like to thank Dr. Ajay Y. Deo for his support during this work.